\documentclass[%
reprint,
superscriptaddress,
amsmath,amssymb,
aps,
pra,
]{revtex4-1}
 
\usepackage{amsmath}
\usepackage{amssymb}
\usepackage{amsfonts}
\usepackage{dsfont}
\usepackage{graphicx}
\usepackage{hyperref}
\usepackage[utf8]{inputenc}
\usepackage[T1]{fontenc}
\usepackage[english]{babel}
\usepackage[dvipsnames]{xcolor}
\usepackage{tikz}
\usepackage{tikz-feynman}
\usepackage{ulem}

\graphicspath{{../figures/}}

\newcommand{\avg}[1]{\left\langle #1 \right\rangle}
\newcommand{\bigexp}[1]{\exp{\left\{ #1 \right\}}}

\let\Re\relax
\DeclareMathOperator{\Re}{Re}
\let\Im\relax
\DeclareMathOperator{\Im}{Im}

\begin{document}
\title{Lemniscate phase trajectories for high-fidelity GHZ state
preparation in trapped-ion chains}
\date{\today}
\author{Evgeny V. Anikin}
\email[]{evgenii.anikin@skoltech.ru}
\affiliation{Russian Quantum Center, Skolkovo, Moscow 143025, Russia}
\author{Andrey Chuchalin}
\affiliation{Russian Quantum Center, Skolkovo, Moscow 143025, Russia}
\affiliation{Moscow Institute of Physics and Technology, Dolgoprudny, 141700, Russia}
\author{Dimitrii Donchenko}
\affiliation{Russian Quantum Center, Skolkovo, Moscow 143025, Russia}
\affiliation{National Research Nuclear University MEPhI, Moscow 115409, Russia}

\author{Olga Lakhmanskaya}
\affiliation{Russian Quantum Center, Skolkovo, Moscow 143025, Russia}
\author{Kirill Lakhmanskiy}
\affiliation{Russian Quantum Center, Skolkovo, Moscow 143025, Russia}

\begin{abstract}
  In trapped-ion chains, multipartite GHZ states can be prepared natively with the help of a single
  bichromatic laser pulse.  However, higher-order terms in the expansion in the Lamb-Dicke parameter $\eta$ limit the
  GHZ state preparation infidelity for rectangular and bell-like pulses to the order of $\eta^4$. 
  For tens of ions, the infidelity caused by out-of-Lamb-Dicke effects can reach several percents.
  We propose an amplitude and phase-modulated pulse shape, an ``echoed lemniscate pulse'', 
  which cancels this contribution into error in the leading order.
  For the proposed pulse, the infidelity scales as $\eta^6$. The improved scaling is achieved because of a special phase
  trajectory of a collective motional mode following the figure-eight curve (lemniscate). We demonstrate
  that the lemniscate pulse allows achieving lower infidelity than bell-like pulses, which can be as low as $10^{-4}$
  for 20-ion chains.
\end{abstract}

\maketitle

\section{Introduction}
The Greenberger-Horne-Zeilinger (GHZ) state is a
protopypical example of a multi-qubit non-classical state
\cite{Greenberger1990, Gisin1998, Frowis2018}.  It has
various applications in quantum information processing
\cite{Gottesman1999}, quantum metrology and sensing
\cite{Giovannetti2006, Marciniak2022}, and quantum
cryptography \cite{Hillery1999}.  Due to high fragility of
the GHZ states to decoherence \cite{Pezze2018},
high-fidelity GHZ state preparation serves as a benchmark
for quantum computing devices \cite{Bao2024}. Although the
achieved GHZ state fidelities grow steadily over last
decades for various quantum computer architectures,
high-fidelity GHZ state
preparation still poses a major challenge for large
quantum registers.




A natural way to prepare $n$-qubit GHZ states exists for
cold trapped ions, which are a well-established platform for
the implementation of a quantum computer featuring long
coherence times, high-fidelity gates, and efficient readout
\cite{Bermudez2017, Bruzewicz2019}.
In trapped-ion chains, GHZ states can be prepared using a
certain $n$-qubit entangling operation, global
M{\o}lmer-S{\o}rensen (MS) gate \cite{Soerensen2000,
Sackett2000, Leibfried2005, Monz2011, Pogorelov2021}.  For
implementing global MS gate, ions are illuminated by a
bichromatic laser beam symmetrically detuned from qubit
transition.  In the Lamb-Dicke regime, the bichromatic beam
creates a qubit-state-dependent force (or spin-dependent
force, SDF) acting on ions \cite{Haljan2005}.  The SDF
entangles qubits with the center-of-mass (COM) phonon mode,
which in turn causes entanglement between qubits. Because of
the equal couplings between qubits and the COM mode, the
resulting evolution operator (global MS gate operator) is
represented by a unitary $\exp{\{-\frac{i\pi}{2}\hat
S_x^2\}}$, where $S_\alpha=\frac12 \sum_i
\hat\sigma_\alpha^{(i)}$  are the collective pseudospin
operators, and $\hat\sigma_{\alpha}^{(i)}$, $\alpha \in
\{x, y, z\}$ is the Pauli sigma matrix operator acting on
the $i$-th ion.  The application of the global MS gate to
the initial $|1\dots 1\rangle$ qubit state results in a
GHZ-like state \cite{Molmer1999}.

However, the infidelity of the resulting GHZ state quickly
grows with the increasing number of ions in the chain and
achieves tens of percents for chains of $\sim 20$ ions
\cite{Monz2011, Pogorelov2021}.  The sources of error are of
the same nature that the errors of the two-qubit MS gate.
One group of errors originates from the presence of
technical noises, such as laser field, magnetic field, and
trapping field fluctuations \cite{Wu2018}. Decoherence
caused by these factors grows quadratically with the number
of ions \cite{Monz2011}.  The other
group of errors originates from the unwanted effects of the
laser-ion interaction. The latter includes the excitation
of higher-frequency (spectator) phonon modes
\cite{Shapira2020}, carrier excitation \cite{Roos2007,
Kirchmair2008, Wu2018, Bazavan2023}, and out-of-Lamb-Dicke
effects \cite{Wu2018, Bluemel2024, OrozcoRuiz2025}. 

The role of the out-of-Lamb-Dicke
effects in the GHZ state preparation is the focus of the
current manuscript.  Using pertubation theory in the
Lamb-Dicke parameter $\eta$, we show that the total error
grows faster than quadratically with the number of ions
and can reach up to $\sim 10$ percent for $20$ ions in the
chain.  On the other hand, it does not decrease with
increasing gate duration, unlike the contribution of the
off-resonant carrier transition and spectator phonon
modes.  Therefore, the out-of-Lamb-Dicke effects pose a
fundamental limitation on the GHZ state fidelity even
under ideal experimental conditions.


To mitigate this error,
we propose a special kind of amplitude and phase modulation of the laser pulse 
shape. 
The pulse shape is designed so that
the phase trajectory of the COM mode follows the figure-eight curve, which is a slight modification of the algebraic curve called the Lemniscate of Gerono \cite{Lawrence2013Catalog}. 
We prove that for the proposed pulse, infidelity scales as $\eta^6$ in contrast with $\eta^4$ for bell-like pulses. 
With numerical simulations, we show that the proposed pulse shape allows achieving GHZ state preparation infidelities of $10^{-6}$-$10^{-5}$ for the chains with up to 20 ions. 
We believe that the proposed pulse configurations can become
a valuable tool for GHZ state preparation in trapped-ion chains
and facilitate various tasks of trapped-ion quantum state engineering.


\section{GHZ state preparation in trapped-ion chains}
\label{sec:setup_description}
 


\begin{figure}
  \includegraphics[width=0.8\linewidth]{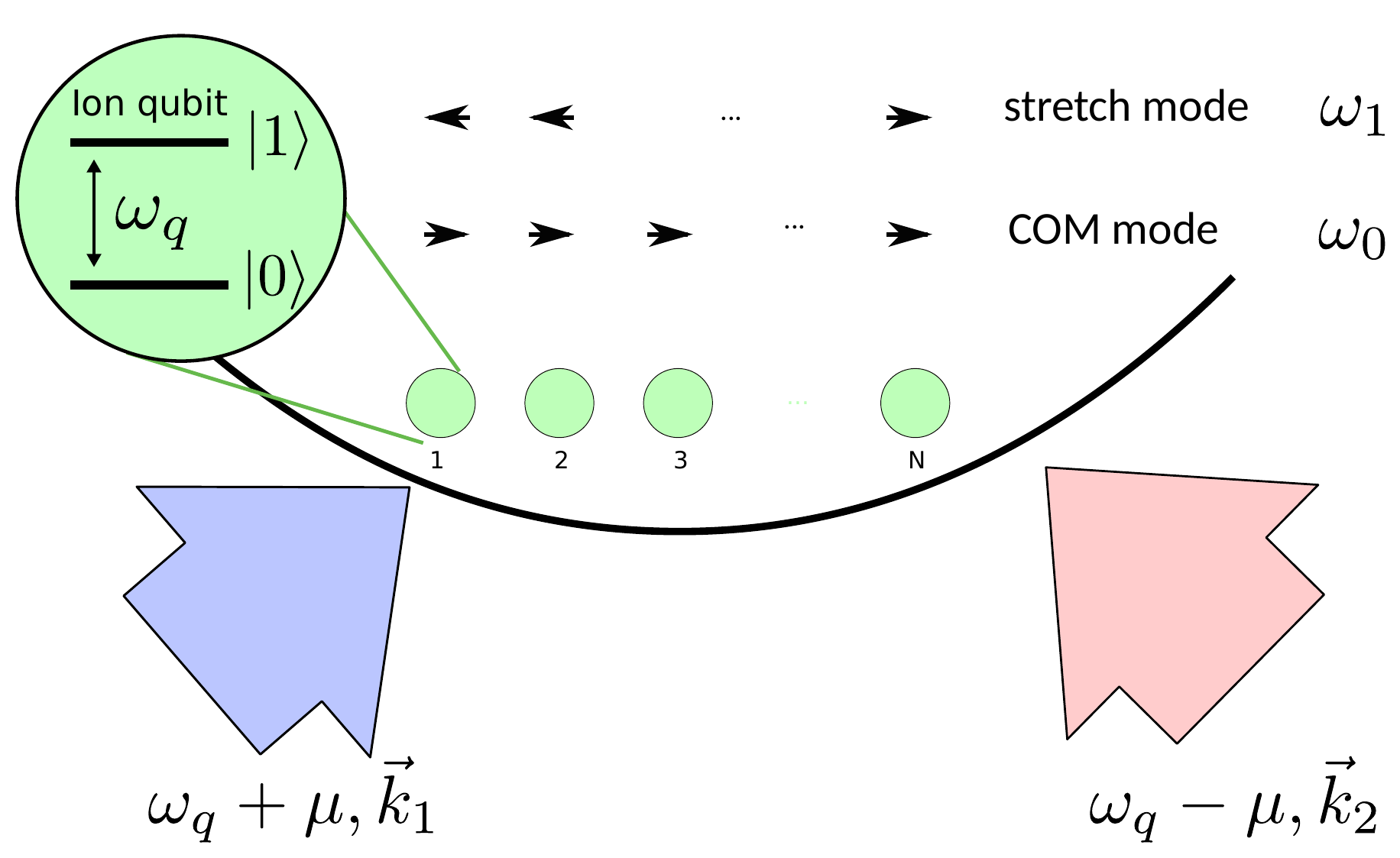}
  \caption{A schematic drawing of a trapped-ion experimental setup.
    An ion chain in a Paul trap is illuminated
    by bichromatic laser field. The field of a Paul trap creates a 
    quadratic pseudopotential confining the ions. The ions form a linear
    ion crystal. Two ion levels with the transition frequency $\omega_q$ 
    form a qubit. The chain is illuminated by two beams with 
    the frequencies $\omega_q \pm \mu$, which cause the interaction between 
    qubit states and chain phonon modes. Two low-frequency axial phonon modes are
    shown: the COM mode with frequency $\omega_0$ and the stretch mode with
    frequency $\omega_1 = \sqrt{3}\omega_0$.
  }
  \label{fig:setup}
\end{figure}

For the analysis of the GHZ state preparation in a
trapped-ion quantum processor, we consider the setup
depicted in  Fig.~\ref{fig:setup}.  A linear ion chain is
confined in a radio-frequency Paul trap along the $z$ axis.
It is illuminated by two laser beam components with the
detunings $\pm \mu$ from the qubit frequency, wavevectors
projections on the chain axis $k_{z1}$ and $k_{z2} = \pm
k_{z1}$ and equal field amplitudes $\Omega(t)$. 
The Hamiltonian for the interaction between the ion qubits, the chain axial motional modes, and the laser beams reads
\begin{equation}
  \begin{gathered}
    \hat{H} = -\frac{i}{2}\sum_i
    \Omega(t)( e^{-i\mu t + ik_{z1} \hat z_i} 
   +  e^{ i\mu t + ik_{z2} \hat z_i}) 
     \sigma_+^{(i)} + \mathrm{h.c.},\\
     \hat{z}_i = \sum_m \sqrt{\frac{\hbar}{2M\omega_m}}  b_{im}
     (\hat a_m e^{-i\omega_m t} + \hat a_m^\dagger e^{i\omega_m t}),
 \end{gathered}
 \label{eq:general_ham}
\end{equation}
where $\hat{z}_i$ are the displacements of the ions along the 
chain, $M$ is the ion mass, $m$ enumerates the chain axial phonon modes,
$\omega_m$ are the frequencies of the phonon modes, $b_{im}$ is the 
matrix of normalized mode vectors, and $a_m, a_m^\dagger$ are the 
annihilation and creation operators for mode $m$.
With different choices of $k_{z1}$ and $k_{z2}$, the Hamiltonian
\eqref{eq:general_ham} can describe different beam geometries. 
For $k_{z1} = k_{z2} = k_z$ (co-propagating beams), the Hamiltonian 
corresponds to the so-called phase sensitive geometry, 
and for $k_{z1} = -k_{z2} = k_z$ (counter-propagating beams), 
it corresponds to the phase insensitive geometry \cite{Lee2005}.

The Hamiltonian \eqref{eq:general_ham} can be simplified using the 
expansion in Lamb-Dicke parameter $\eta_{im}$ and the rotating wave approximation. 
\begin{equation}
  \eta_{im} = k_z\sqrt{\frac{\hbar}{2M\omega_m}} b_{im}.
  \label{eq:ld_param_def}
\end{equation}
First, the exponents $e^{ik_{z1}\hat{z}_i}$, $e^{ik_{z2}\hat{z}_i}$ are
expanded in power series: 
$e^{ik_{z1}\hat{z}_i} = 1 + ik_{z1}\hat{z}_i + \dots$, where the small parameters of the expansion are the Lamb-Dicke parameters 

We assume, that the detuning $\mu$ is close to the COM mode
frequency $\omega_0$: $\mu = \omega_0 + \delta$. For the
rotating-wave approximation (RWA), one needs to keep only
terms oscillating with frequencies $\sim\delta$ and neglect
the terms oscillating with frequencies $\sim\omega_0$. After
RWA, only the odd terms of the Lamb-Dicke expansion and only
the contribution of the COM mode survive. In particular, the
fast oscillating carrier terms coming from the zero-order
terms of the expansion of the exponent are neglected.
Therefore, the leading-order contribution in the Lamb-Dicke
expansion comes from the linear terms $k_{z1}\hat{z}_i$,
$k_{z2}\hat{z}_i$.  This results in an SDF Hamiltonian
\begin{equation}
  \hat{H} = \eta(\Omega^*(t)\hat ae^{-i\delta t}  
  + \Omega(t)\hat a^\dagger e^{i\delta t} ) \hat S_x,
  \label{eq:sdf_ham}
\end{equation}
where $\eta = \sqrt{\frac{\hbar}{2M\omega_0n}}$ 
is the Lamb-Dicke parameter of the COM mode.

The Hamiltonian \eqref{eq:sdf_ham} can be solved exactly. Its evolution operator reads
\begin{equation}
\hat{U}\left( t_{2},t_{1} \right) = \exp\left( - 2i\chi\left( t_{2},t_{1} \right)S_{x}^{2} \right)D\left( 2\alpha\left( t_{2},t_{1} \right)S_{x} \right),
\label{eq:ms_evo_op}
\end{equation}
Here \(S_{x} = \frac{1}{2}\sum_{i}^{}\sigma_{x}^{i}\) is the collective spin operator, and the quantities
\(\alpha\left( t_{2},t_{1} \right)\) and 
\(\chi\left( t_{2},t_{1} \right)\) are calculated as 
\begin{equation}
    \alpha(t_2, t_1) = -\frac{i\eta}{2}\int_{t_1}^{t_2} dt \,\Omega(t) e^{-i\delta t} ,\\
    \label{eq:alpha_sdf}
\end{equation}
\begin{multline}
    \chi\left( t_{2},t_{1} \right) = \eta\Re\int_{t_{1}}^{t_{2}}dt\,\Omega^*\left( t \right)e^{i\delta t}\alpha\left( t,t_{1} \right)= \\= -i\int \alpha d\alpha^* - \alpha^* d\alpha.
    \label{eq:chi_sdf}
\end{multline}
The quantity \(\alpha\left( t_{2},t_{1} \right)\) has the
meaning of the displacement amplitude of the phonon mode,
and \(\chi\left( t_{2},t_{1} \right)\) represents the
spin-spin coupling phase.  If the pulse parameters are
chosen so that $\alpha = 0$ at the end of the pulse, the
evolution operator reduces to the required global MS gate
operator:
\begin{equation}
  \hat{U} = \bigexp{-2i\chi S_x^2}.
\end{equation}
The conditions for the GHZ state implementation take the form
\begin{equation}
    \begin{gathered}
        \alpha(t_f, t_0) = 0,\\
        \chi(t_f, t_0) = \pi/4,
    \end{gathered}
    \label{eq:ideal_gate_conditions}
\end{equation}
where \(t_{0}\) and \(t_{f}\) refer to the beginning and the end of the laser pulse.

As $\alpha = 0$ at the end of the pulse, the amplitude $\alpha(t, t_0)$ follows a closed trajectory in the complex plane. 
According to Eq.~\eqref{eq:chi_sdf}, the spin-spin couping phase $\chi$ equals four times the oriented area of the region 
enclosed by the trajectory.

\begin{figure*}
    \includegraphics[width=\linewidth]{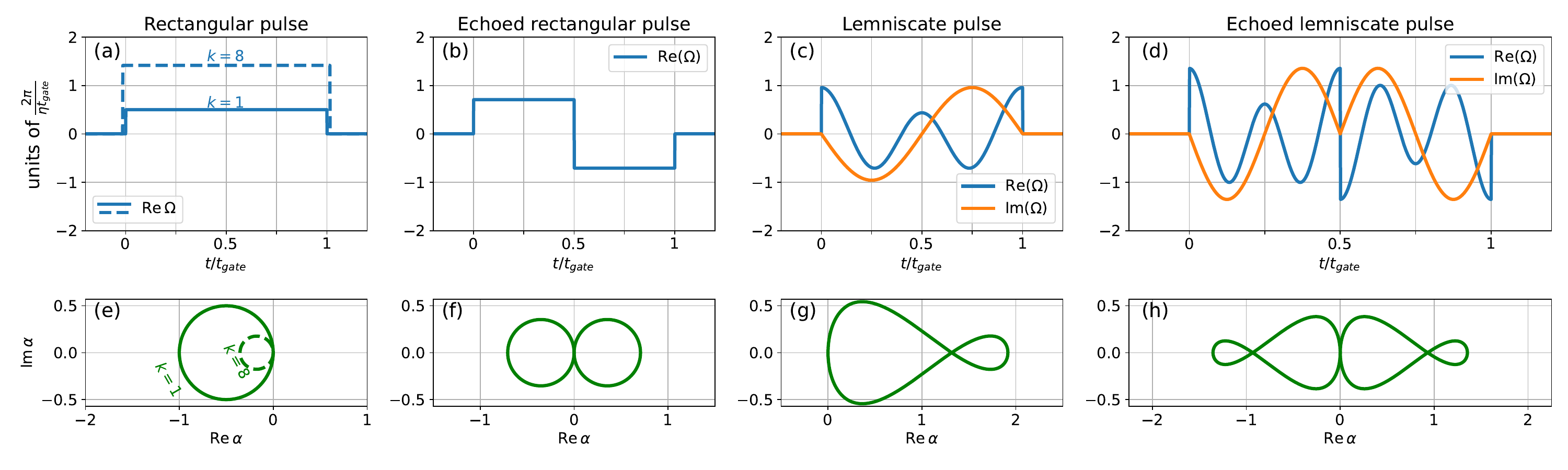}
    \caption{Pulse shapes and phase trajectories for (a, e) rectangular pulse, (b, f) echoed rectangular pulse, (c, g) lemniscate pulse, (d, h) echoed lemniscate pulse.
  }
    \label{fig:phase_trajectories}
\end{figure*}

For the rectangular pulse
\(\Omega_{0}\left(t \right) = \text{const}\), the integrals for  \(\alpha\) and \(\chi\)
can be calculated analytically: this results in 
\begin{equation}
  \alpha\left( t_{2},t_{1} \right) = \frac{\eta\Omega_{0}e^{- i\psi}}{2\delta}\left( e^{- i\delta t_{2}} - e^{- i\delta t_{1}} \right),
  \label{eq:alpha_rect}
\end{equation}
\begin{equation}
  \chi(t_2, t_1) = \frac{\eta^2\Omega^2}{2\delta}
  \left(
    t_2 - t_1 - \frac{\sin{\delta(t_2 - t_1)}}{\delta}
  \right).
  \label{eq:chi_rect}
\end{equation}
The global MS gate implementation conditions \eqref{eq:ideal_gate_conditions} 
for the rectangular pulses
lead to the following expressions for $\delta$ and $\Omega_0$:
\begin{equation}
\delta = \frac{2\pi k}{t_{gate}}
\label{eq:ms_cond_delta}
\end{equation} 
where \(t_{\text{gate}} = t_{f} - t_{0}\), and 
\begin{equation}
  \Omega_{0} = \frac{\pi\sqrt{k}}{\eta t_{gate}}. 
\label{eq:ms_cond_Omega}
\end{equation}
At these parameters, the phase trajectory $\alpha(t, t_0)$ consists of $k$ circles of the radius $\sqrt{k}/4$ (see Fig.~\ref{fig:phase_trajectories}a,e). 

For non-rectangular pulses, the integrals for $\alpha$ 
and $\chi$ should be calculated either analytically or numerically, and 
Eqs.~\eqref{eq:ideal_gate_conditions} should be solved to find pulse
parameters.


Let us discuss the conditions which are necesary for the validity of the
approximations used to derive the effective SDF 
Hamiltonian \eqref{eq:sdf_ham}.

The single-mode approximation is valid  when
the excitation of all modes except the COM remains negligible. The
excitation amplitude of the mode $m$ with frequency $\omega_m$ can be 
estimated as $\sim\eta\Omega/(\mu - \omega_m) \approx 
\eta\Omega/(\omega_0 - \omega_m)$. 
Global MS gate implementation conditions \eqref{eq:ideal_gate_conditions} require that 
$\eta\Omega \sim \delta$, so the excitation amplitude becomes 
$\sim\delta/(\omega_0 - \omega_m) \sim 2\pi/(t_{gate}(\omega_0 - \omega_m))$. 
Therefore, the single-mode
approximation is valid providing that the gate time satisfies $t_{gate} \gg (\omega_0 - \omega_m)^{-1}$. 
Among all axial modes, the breathing 
or stretch mode has the closest frequency to the COM mode, and it 
equals $\sqrt{3}\omega_0$ \cite{James1998}. 
As a result, the single-mode approximation requires 
that $\omega_0t_{gate} \gg 1$.

With increasing number of ions in the chain, 
$\omega_0$ should be lowered to maintain 
the chain stability. However, the condition $\omega_0 t_{gate} \gg 1$
could still be well satisfied for $\sim 20$ ions. For example, the typical values of the 
COM mode frequency for chains of $\sim 20$ ${}^{40}\mathrm{Ca}^+$ ions 
can be $100-200$ kHz (see Appendix~\ref{appendix:ion_chains}).  
So, the gate time of hundreds of microseconds is enough for the single-mode 
approximation. 
These parameters are typical for up-to-date ion trap experimens, so we assume
them in the further consideration.

For the validity of RWA, the same arguments apply,
because RWA neglects the terms oscillating with frequencies $\sim\omega_0$.

Now let us analyze the effect of the high-order terms in the Lamb-Dicke 
expansion \cite{Vogel1995}. For that, we will apply RWA to the Hamiltonian 
\eqref{eq:general_ham} while keeping all orders in the Lamb-Dicke expansion.
As shown in Appendix~\ref{appendix:all_ord_rwa_ham},
the approximate Hamiltonian takes the same form for phase-sensitive
and phase-insensitive geometries:
\begin{equation}
  \begin{gathered}
    \hat{H} = \eta \sum_{n} (\Omega^*(t)\hat{A}e^{-i\delta t} + 
     \Omega(t)\hat{A}^\dagger e^{i\delta t})S_x,\\
     \hat{A} = \frac{1}{i\eta}\sum  
     \langle n | e^{i\eta(a + a^\dagger)} | n+1\rangle
     |n\rangle\langle n+1|.
  \end{gathered}
  \label{eq:all_ord_rwa_ham}
\end{equation}
This Hamiltonian has a form similar to the SDF Hamiltonian \eqref{eq:sdf_ham},
however, instead of the annihilation and creation operators 
$\hat a, \hat a^\dagger$, it contains the modified ladder operators 
$\hat{A}, \hat{A}^\dagger$. The operators $\hat{A}$ can be expressed 
via $\hat{a}$ through the power series in $\eta$
\begin{equation}
  \hat{A} = (1 - \eta^2)\hat{a} - 
  \frac{\eta^2}{2}\hat{a}^\dagger \hat{a} \hat{a} + \dots,
  \label{eq:A_expansion}
\end{equation}
and $\hat{A} \to \hat{a}$ in the limit $\eta \to 0$.

The Hamiltonian \eqref{eq:all_ord_rwa_ham} is invariant with respect to 
simultaneous rescaling of $t$, $\delta$ and $\Omega(t)$:
$t\to \lambda t$, $\delta \to \lambda^{-1}\delta$, 
$\Omega(t) \to \lambda^{-1}\Omega(\lambda t)$. Therefore, the
results concerning state preparation and gate fidelities obtained
from this Hamiltonian does not depend on the choice of $t_{gate}$
if the performed approximations hold.


The typical value of $\eta$ for the COM mode in trapped-ion chains 
is of order $0.03$-$0.05$ (see the Appendix~\ref{appendix:ion_chains}). 
From \eqref{eq:A_expansion}, one should expect that the contribution
of the high-order terms in the Lamb-Dicke expansion scales as $\eta^4$.
For small number of ions, this would be a negligible contribution. However,
as will be shown below, the effect of the higher-order terms
grows rapidly with the increasing number of ions and can become 
a dominant contribution under realistic experimental conditions.

In the next sections, we analyze the contribution of the high-order terms
in the Lamb-Dicke expansion into GHZ state preparation infidelity and 
propose a method to minimize this contribution.





\section{The effects of the high-order terms in Lamb-Dicke expansion}
\label{sec:high_ord_effects}
For the analytical treatment of the leading-order contribution of 
the out-of-Lamb-Dicke effects, we will further simplify the 
Hamiltonian \eqref{eq:all_ord_rwa_ham} and consider 
the expansion only up to the third order in $\eta$:
\begin{equation}
  \hat{H} = \eta\Omega^*(t)e^{i\delta t}\left[\hat{a}\left(1 - \frac{\eta^2}{2}\right) - 
  \frac{\eta^2}{2}\hat{a}^\dagger \hat{a}\hat{a}\right]\hat S_x + \mathrm{h.c.}
    \label{eq:ham_3ord}
\end{equation}
We find the evolution operator for \eqref{eq:ham_3ord} using 
perturbation theory with the 3-order terms in $\hat{a}, \hat a^\dagger$
considered as a perturbation. To simplify the expressions, we
renormalize $\Omega(t)$ as follows:
\begin{equation}
  \Omega'(t) = \left(1 - \frac{\eta^2}{2}\right)\Omega(t).
  \label{eq:Omega_renorm}
\end{equation}
Then, only the term proportional to $\hat a^\dagger \hat a \hat a$ 
remains as a perturbation. We take the zero-order evolution operator
from Eq.~\eqref{eq:ms_evo_op}, where $\alpha(t)$ and $\chi(t)$ 
are calculated using the renormalized field amplitude 
$\Omega'(t)$. In the interaction picture, the annihilation operator
becomes,
\begin{equation}
  U_0^\dagger\hat{a}\hat{U} = \hat{a} + 2\alpha \hat \hat S_x, 
\end{equation}
so the interaction picture Hamiltonian reads
\begin{equation}
  \hat{V} = -\frac{\eta^3\Omega'^*(t)}{2}e^{i\delta t}(\hat a^\dagger + 2\alpha^*(t) \hat S_x)(\hat a + 2\alpha(t) \hat S_x)^2\hat S_x
  + \mathrm{h.c.}.
  \label{eq:int_ham}
\end{equation}
Then, we find the evolution operator with Magnus expansion
(see the derivation in Appendix~\ref{appendix:magnus}). 
In the leading 
order in $\eta$,
\begin{equation}
\hat{U}_I \approx \bigexp{-i\int \hat{V} dt}
= 1 - i\hat{T},
\end{equation}
\begin{equation}
  \hat{T} \approx \delta\theta_2 \hat S_x^2 + 
  \theta_4 \hat S_x^4 + (g \hat{a}^\dagger + g^*\hat{a}) \hat S_x^3 + \dots
  \label{eq:magnus_3ord_op}
\end{equation}
where dots stand for the terms with two or more 
creation and annihilation operators which do not contribute into GHZ preparation error under our 
assumptions (see the discussion in Appendix~\ref{appendix:magnus}).
We included the term $\delta\theta_2$ 
responsible for the deviation $\Delta\Omega'$ of the field amplitude
from the leading-order optimal value defined by \eqref{eq:ms_cond_Omega}:
\begin{equation}
  \delta\theta_2 = \frac{\pi\Delta\Omega'}{\Omega'}.
\end{equation}
The coefficients $\theta_4$, $g$, can be
found from the expressions
\begin{equation}
    \theta_4 = 8i\eta^2 \int |\alpha^2|(\alpha d\alpha^* - \alpha^* d\alpha),
    \label{eq:theta_4}
\end{equation}
\begin{equation}
    g =  4\eta^2\int (\alpha^2d\alpha^* - 2|\alpha|^2d\alpha).
    \label{eq:ph_ex_ampl}
\end{equation}

For rectangular pulses with $k$ circles, $\theta_4$ and $g$ 
can be calculated analytically using Eqs.~\eqref{eq:theta_4},
\eqref{eq:ph_ex} and the expressions \eqref{eq:alpha_rect}, \eqref{eq:chi_rect} for 
$\alpha$ and $\chi$:
\begin{equation}
  \theta_4 = -\frac{3\pi\eta^2}{8k},
  \label{eq:theta_4_rect}
\end{equation}
\begin{equation}
  g = \frac{i\pi\eta^2}{2\sqrt{k}}
  \label{eq:g_rect}.
\end{equation}
For other bell-like pulse shapes, $\theta_4$ and $g$ are also of the order
$O(\eta^2)$.

Let us discuss the errors originating from Eq.~\eqref{eq:magnus_3ord_op}. 
Two main effects of the high-order corrections to the evolution operator can be identified.
First, there is a term which corresponds to the phonon creation at the 
end of the gate operation: the probability of the 
phonon creation can be calculated as 
\begin{equation}
    P_\mathrm{ph} = |g|^2\avg {\hat S_x^6}.
    \label{eq:ph_ex}
\end{equation}
Second, there is a term proportional to $\hat S_x^4$ which modifies the action of the evolution operator in the qubit subspace
and causes the deviation of the final state from the target GHZ state. 
These effects are the leading-order perturbation theory contributions 
into the infidelity of the GHZ state.




Assume that the initial qubit state is
$|\psi_0\rangle = |1\dots 1\rangle$, and the phonon mode is
prepared in the ground state. Then, the GHZ state infidelity 
can be calculated from the evolution operator in Eq.~\eqref{eq:magnus_3ord_op}: 
\begin{multline}
  1-F = \langle\psi_0|\hat T^\dagger \hat T|\psi_0\rangle - \langle\psi_0|\hat T^\dagger|\psi_0\rangle\langle\psi_0|\hat T|\psi_0\rangle 
  \\ = \theta_4^2(\avg{S_x^8} - \avg{S_x^4}^2) 
  + 2\delta\theta_2\theta_4 (\avg{S_x^6} - \avg{S_x^4}\avg{S_x^2})
  \\+ \delta\theta_2^2(\avg{S_x^4} - \avg{S_x^2}^2) + P_\mathrm{ph}.
  \label{eq:inf_w_sx4}
\end{multline}
where $\avg{S_x^n}$ denote the averages over $|\psi_0\rangle$.
As stated above, the total error contains two contributions:
the part originating from the qubit subspace (the 
$\hat S_x^4$ term) 
and the part originating from phonon excitation. 

The error \eqref{eq:inf_w_sx4} can be reduced by an 
adjustment of the pulse amplitude, which is regulated by the 
previously introduced parameter $\delta\theta_2$. 
As Eq.~\eqref{eq:inf_w_sx4} is a quadratic function
of $\delta\theta_2$, the optimal value of 
$\delta\theta_2$ can be easily found:
\begin{equation}
  \delta\theta_2 = -\frac{(\avg{S_x^6} - \avg{S_x^4}\avg{S_x^2})}{(\avg{S_x^4} - \avg{S_x^2}^2)}\theta_4, 
\end{equation}
The optimal value of $\Delta\Omega$ reads
\begin{equation}
  \Delta\Omega' = \left(\frac{\delta\theta_2}{\pi} + \frac{\eta^2}{2}\right)\Omega_0,
  \label{eq:delta_omega_rect_pulse_theory}
\end{equation}
\typeout{Rectangular pulse amplitude correction: Eq. \theequation}
where the additional $\eta^2/2$ term arises due to the renormalization
\eqref{eq:Omega_renorm}.
At the optimal amplitude value,
\begin{multline}
    1-F = \theta_4^2\left(
    \avg{S_x^8} - \avg{S_x^4}^2
    - \frac{(\avg{S_x^6} - \avg{S_x^4}\avg{S_x^2})^2}{\avg{S_x^4} - \avg{S_x^2}^2}
  \right) \\+ |g^2|\avg{S_x^6}.
  \label{eq:inf_eta4}
\end{multline}
\typeout{Rectangular pulse optimal infidelity: Eq. \theequation}
The large-$n$ asymptotics of the $\hat{S}_x^4$ contribution 
and the phonon creation probability 
(the coefficients at $\theta_4$ and $|g|^2$)
are $n^4$ and $n^3$ respectively.

\begin{figure}
  \includegraphics[width=\linewidth]{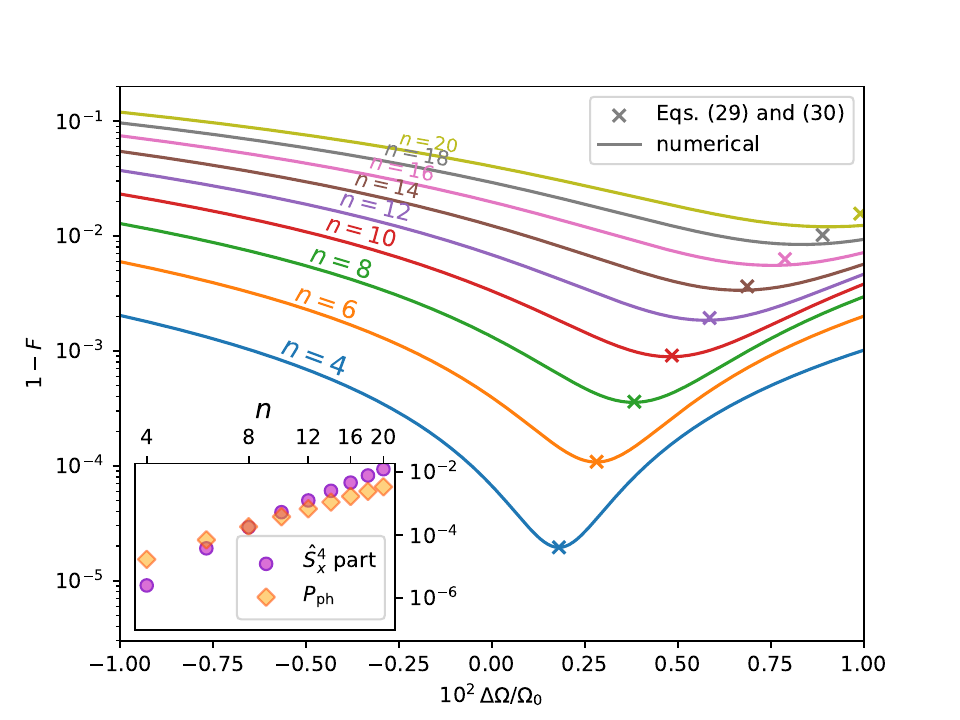}
  \caption{The GHZ state preparation infidelity as a 
  function of field amplitude of a rectangular pulse for the number of ions from 4 to 20.
  Solid lines indicate the numerical results, and
  crosses indicate the optimal amplitude values and infidelities calculated from Eq.~\eqref{eq:delta_omega_rect_pulse_theory} and \eqref{eq:inf_eta4}.
The inset: the analytically calculated $S_x^4$ and phonon contributions into GHZ error as functions of $n$.
}
  \label{fig:rect_pulse_n_ions_dep}
\end{figure}

In Fig.~\ref{fig:rect_pulse_n_ions_dep}, 
we present the infidelity values for the optimal field amplitudes
calculated from Eqs.~\eqref{eq:delta_omega_rect_pulse_theory} and \eqref{eq:inf_eta4} together with the
results of the numerical simulation of the Hamiltonian~\eqref{eq:all_ord_rwa_ham} for the Lamb-Dicke 
parameter $\eta = 0.03$. We solve the time-dependent
Schr{\o}dinger equation (TDSE) with the 
initial state $|0\dots 0\rangle$ for rectangular 
pulse $\Omega(t)$ 
with the detuning $2\pi/t_{gate}$ and the amplitude close to $\frac{\pi}{\eta t_{gate}}$ (this corresponds 
to $k=1$ circles). For different numbers of ions, we calculate the GHZ infidelity as a function of the field amplitude. 
A discrepancy between the analytical and numerical values of $\Delta\Omega/\Omega$ and $1-F$ comes from the 
high-order expansion terms $\sim\eta^6$ present in the Hamiltonian ~\eqref{eq:all_ord_rwa_ham} and 
the higher-order terms of the perturbation theory.

According to the results of Fig.~\ref{fig:rect_pulse_n_ions_dep}, the considered contributions 
into error grow faster than quadratically with the number of ions
For $\sim 20$ ions, the error becomes larger than $10^{-2}$, which 
is comparable to the up-to-date level of technical noise contributions. 
From that, we conclude that the construction
of the pulses minimizing the high-order contribution is desirable to achieve high GHZ state fidelities.

\section{Approaches for error mitigation}
\label{sec:lemn_pulse}
In this section, 
we discuss the approaches for laser pulse design 
to minimize the infidelity caused by out-of-Lamb-Dicke effects.

As can be seen 
from Eqs.\eqref{eq:theta_4_rect}, \eqref{eq:g_rect} and
\eqref{eq:inf_eta4},
it is possible to decrease the 
contribution of high-order terms by using pulses with 
$k > 1$, as $\theta_4$ and 
$g$ decrease with increasing $k$.
It is also possible to design pulses for which
$\theta_4$ and $g$ vanish, which
is the goal of the current manuscript. 

The contribution of $g$ can be canceled 
as follows.
For any pulse shape $\Omega(t)$, it is possible to construct 
an echoed version of the pulse which cancels the leading-order contribution
into phonon creation. We define it
as
\begin{equation}
    \Omega_\mathrm{echoed}(t) = 
    \begin{cases}
      \sqrt{2}\Omega\left(2t\right), & t < t_{gate}/2,\\
      -\sqrt{2}\Omega\left(2(t-t_{gate})\right), & t > t_{gate}/2.
    \end{cases}
\end{equation}
The echoed version of the pulse has the same duration and twice the intensity and the detuning as the original pulse. 
If $\Omega(t)$ satisfies the gate conditions 
\eqref{eq:ideal_gate_conditions}, the same holds for 
the echoed version.
Due to the 
symmetry of the echoed pulse, 
$g = 0$. As an example, we show 
the echoed rectangular pulse
and the corresponding phase 
trajectory in Fig.~\ref{fig:phase_trajectories}(b,f).

In the next subsection, we will construct the 
pulse which satisfies the condition $\theta_4 = 0$. 
For its echoed version, both of the contributions 
into error of the order $\eta^4$ cancel,
so the total error for the pulse 
scales as $\eta^6$.



\subsection{Lemniscate pulse construction}

The expression \eqref{eq:theta_4}
for $\theta_4$ can be interpreted as 
the weighted area integral
where each infinitesimal area element is assigned with 
the weight $|\alpha|^2$.
This allows constructing a pulse 
with $\theta_4 = 0$. 
The phase trajectory 
$\alpha(t)$ should form the figure-eight curve starting 
from zero and following 
two loops, clock-wise and counter-clock-wise, 
on the phase plane (see Fig.~\ref{fig:phase_trajectories}(g)). 
The shape and 
size of the loops should be adjusted so that the contributions of the loops
into $\theta_4$ cancel each other while the spin-spin entangling phase $\chi$
remains nonzero. The pulse amplitude $\Omega(t)$ can be found using
Eq.~\eqref{eq:alpha_sdf} by differentiating 
$\alpha(t)$.

We parametrize the figure-eight trajectory 
by the following equation:
\begin{equation}
  \begin{gathered}
    \Re\alpha(t) = A(1 - \cos{\gamma t}),\\
    \Im\alpha(t) = A\sin(\gamma t)(1 - a + a\cos{\gamma t}),
  \end{gathered}
  \label{eq:lemniscate_parametrization}
\end{equation}
where $\gamma = 2\pi/t_{gate}$, and the parameters $a$ and $A$ 
define the shape and the size of the curve.
For $a > 1/2$, the trajectory is a 
figure-eight curve. It reduces to the Lemniscate 
of Gerono  at $a = 1$ \cite{Lawrence2013Catalog}.

The values of $\theta_2$ and $\theta_4$ for this phase trajectory can be 
calculated analytically:
\begin{equation}
  \begin{gathered}
    \chi = \pi A(1 - a),\\
    \theta_4 = \pi A\left(6 - 10a + \frac{7}{2}a^2 - \frac{3}{2}a^3\right).
  \end{gathered}
\end{equation}
The solution of the equations $\chi = \pi/4$, $\theta_4 = 0$ is as follows: 
\begin{equation}
  \begin{gathered}
    a = a_0 \equiv 0.7274789, \\
    A = A_0 \equiv 0.95778915.
  \end{gathered}
  \label{eq:lemniscate_gate_conditions}
\end{equation}
The phase trajectory for these values 
of $a$ and $A$ is an asymmetric figure-eight 
curve shown in 
Fig.~\ref{fig:phase_trajectories}(h). 
Because of the weight $|\alpha|^2$, the smaller
loop of the curve exactly balances the larger
loop in the $\theta_4$ integral. 
Conversely, the contributions of two loops do 
not cancel in the spin-spin entangling phase $\chi$, which allows the 
GHZ state preparation.

For a phase-modulated pulse, we can use
the value of the bichromatic detuning  
$\delta=0$ without loss of generality, which will be taken for further analysis.
Then,we can find $\Omega(t)$ from Eq.~\eqref{eq:alpha_sdf}:
\begin{equation}
  \Omega(t) = i\frac{d\alpha}{dt} = 
-A\gamma[e^{-i\gamma t} - a(\cos{\gamma t} - \cos{2\gamma t})].
\label{eq:lemniscate_pulse}
\end{equation}
Below, we will call this pulse shape a lemniscate pulse.
According to Eq.~\eqref{eq:lemniscate_pulse},
it is both amplitude and phase modulated. The lemniscate 
pulse shape $\Omega(t)$ is shown in Fig.~\ref{fig:phase_trajectories}(d). 

By choosing other parametrizations and shapes for figure-eight curves, 
it is possible to design other pulses satisfying the properties 
$\chi = \pi/4$, $\theta_4 = 0$.



\begin{figure}
  \includegraphics[width=\linewidth]{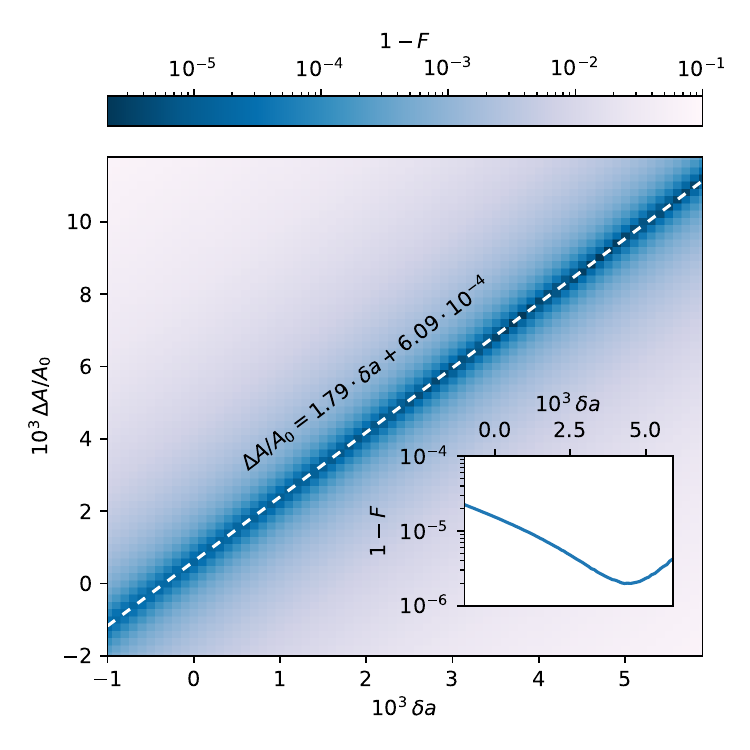}
  \caption{GHZ preparation infidelity for the echoed lemniscate pulse 
  obtained from numerical simulation
  at $\eta = 0.03$ for 20 ions as a 
  function of two parameters $a$, $A$ (see Eq.~\eqref{eq:lemniscate_parametrization}). 
  Here $\delta a = a - a_0$,
$\Delta A = A - A_0$, where $a_0$ and $A_0$ are defined by \eqref{eq:lemniscate_gate_conditions}. Along the dashed line, infidelity is of order $\sim 10^{-5}$. 
The inset: infidelity is plot
as a function of $\delta a$ along the dashed line indicated in (a). The minimum is at 
$\delta a \approx 4.3\cdot 10^{-4}$, 
$\Delta A/A_0 \approx 0.0083$.
}
  \label{fig:lemniscate_2d_fidelity_dep}
\end{figure}

\subsection{Numerical fidelity analysis}
In this subsection, we use the solution of the TDSE with the 
Hamiltonian~\eqref{eq:all_ord_rwa_ham} to compare 
$1-F$ for the lemniscate pulse and $1-F$ for 
the rectangular and the echoed rectangular pulses 
with different values of $k$. We consider different numbers of ions 
and different values of the Lamb-Dicke parameter.


To find the parameters of the lemniscate pulse ensuring optimal 
infidelity, we scan numerically the echoed lemniscate pulse 
parameters $a, A$ in the close vicinity of the values 
\eqref{eq:lemniscate_gate_conditions}. This is necessary because 
the higher-order terms in the Lamb-Dicke 
expansion alter the optimal lemniscate pulse parameters 
\eqref{eq:lemniscate_gate_conditions} by the terms of 
$\sim\eta^2$, similar to the case of the
rectangular pulse analyzed in Section~\ref{sec:high_ord_effects}. 
In Fig.~\ref{fig:lemniscate_2d_fidelity_dep}, 
we show the infidelity of the 20-ion GHZ state
at $\eta = 0.03$ for the echoed lemniscate pulse as 
a function of $\delta a, \Delta A$, where $\delta a = a - a_0$, $\Delta A = A - A_0$. 
According to the results of Fig.~\ref{fig:lemniscate_2d_fidelity_dep},
the minimal infidelity of $\sim 2\cdot 10^{-6}$ is 
achieved at $\delta a \sim 4.3\cdot 10^{-3}$, 
$\Delta A \sim 0.0083$. For rectangular and echoed rectangular pulses,
we use the same procedure: namely, we scan the pulse amplitude in 
the vicinity of the values given by \eqref{eq:ms_cond_Omega}.

\begin{figure}
  \includegraphics[width=\linewidth]{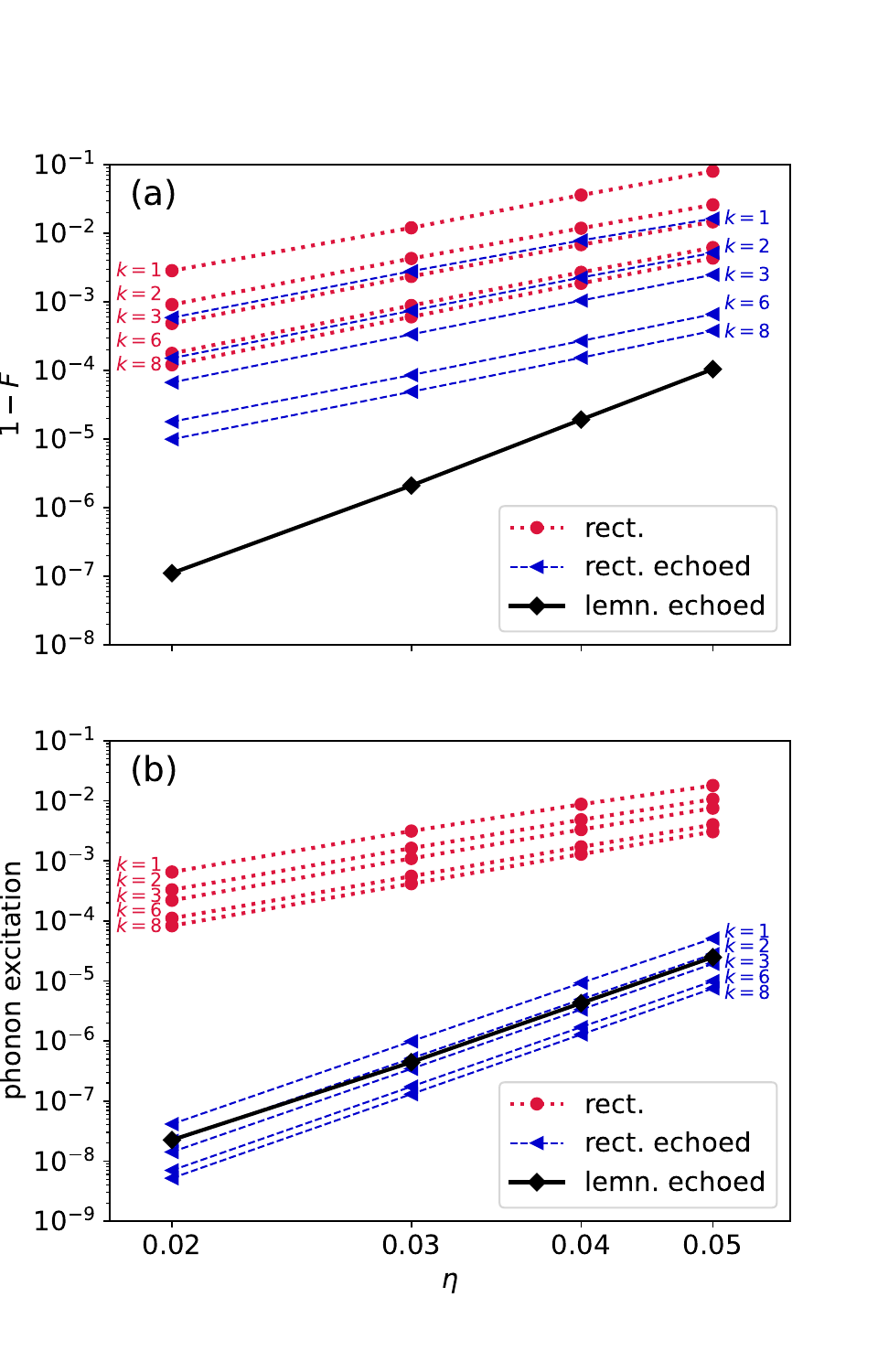}
  \caption{Optimized 20-qubit GHZ state preparation (a) infidelity and 
   (b) phonon excitation probabilities obtained from numerical simulation for 
  rectangular, echoed rectangular, and echoed lemniscate pulse
  as functions of $\eta$.}
  \label{fig:eta_dep_pulse_comparison}
\end{figure}

In Fig.~\ref{fig:eta_dep_pulse_comparison}a, we show the dependencies 
of the GHZ state fidelity for a 20-ion chain as a function of the 
Lamb-Dicke parameter $\eta$ for rectangular, echoed rectangular, and echoed 
lemniscate pulses. For rectangular and echoed rectangular pulses, we consider 
different values of the parameter $k$ (the number of circles) from 1 to 8. 
In agreement with the analytical predictions, infidelity scales as $\eta^4$ 
for (echoed) rectangular pulses and as $\eta^6$ for echoed lemniscate pulse. 
Also, as expected, the infidelity is lower for larger values of $k$. 
However, even for the largest considered
value of $\eta$, $\eta = 0.05$, the infidelity for the lemniscate pulse is lower 
than for the rectangular pulses even at $k=8$. 
 
\begin{figure}
  \includegraphics[width=\linewidth]{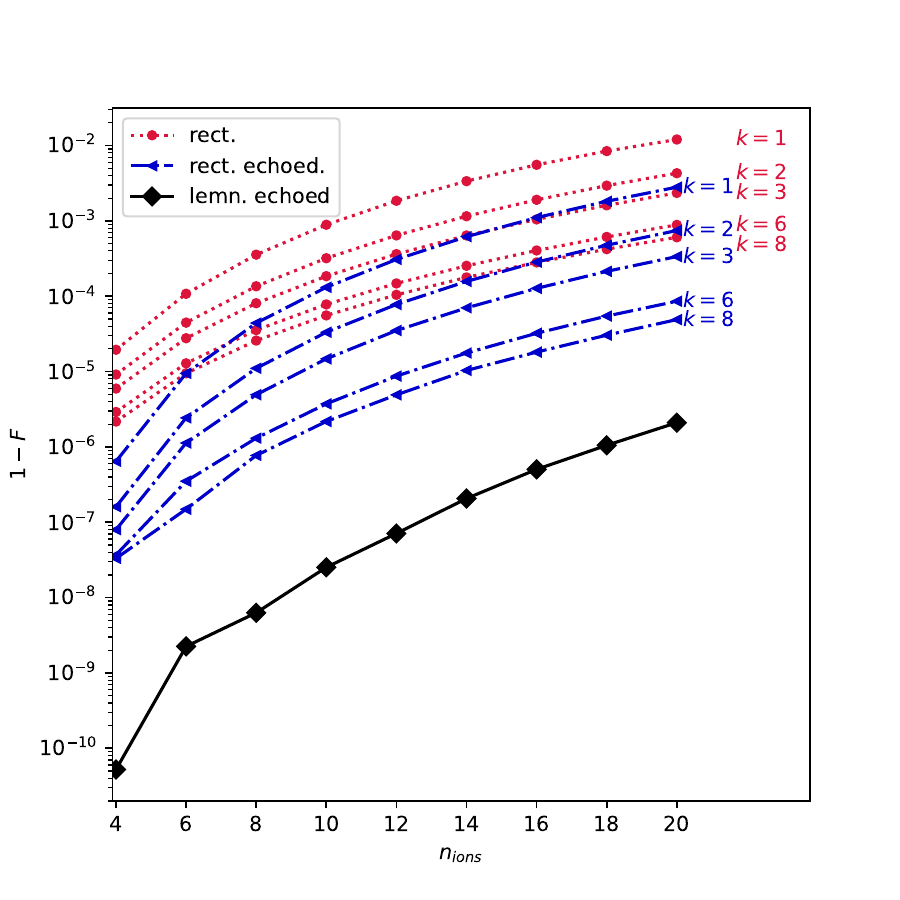}
  \caption{Optimized GHZ state preparation infidelity obtained from numerical 
   simulation at $\eta = 0.03$ 
   as a function of the number of ions for rectangular, echoed rectangular,
  lemniscate, and echoed lemniscate pulses.}
  \label{fig:n_ions_dep}
\end{figure}

In Fig.~\ref{fig:n_ions_dep}, we show the        
dependence of the optimized infidelity 
as a function of the number of ions 
with fixed $\eta = 0.03$ for the 
same types of pulses. It can be seen that the
echoed lemniscate pulse shows significant advantage 
over (echoed) rectangular pulses for all $n < 20$.

\subsection{Discussion}
The advantage of the lemniscate pulse comes with a cost 
of higher required field amplitude or pulse duration. 
Indeed, among all pulses considered in the 
previous section, the rectangular pulse with $k=1$ has the smallest field 
amplitude at a given gate time. All the other pulses, i.e. 
rectangular pulses with 
larger values of $k$, echoed rectangular pulses, and lemniscate pulse, require larger field 
amplitude or/and larger gate time (as the gate time is inversely proportional to 
the field amplitude). As can be seen from Fig.~\ref{fig:phase_trajectories}, the
maximum amplitude of the lemniscate pulse is 
comparable to that of the rectangular pulse
at $k=8$, which is approximately $2.8$ larger than
for the  case $k=1$.
At the same time, the infidelity 
of the lemniscate pulse is considerably lower:
for 20 ions and $\eta \in [0.02, 0.05]$, it ranges 
from $10^{-7}$ to $10^{-4}$, whereas for rectangular 
pulse it ranges from $10^{-4}$ to $10^{-1}$.
So, we conclude that the lemniscate pulse is more suitable for high-fidelity GHZ 
state preparation.

Also, let us discuss the applicability of our results beyond RWA.
The pulses considered in the previous sections (rectangular, echoed rectangular, and echoed
lemniscate) contain discontinuities in the beginning, in the middle, 
and at the end of the gate, which can lead to the errors induced by 
the non-RWA effects \cite{Kirchmair2008}. Therefore, 
the practical performance of the lemniscate pulse could be further 
improved by adding smooth slopes at the discontinuities. Although 
the exact parametrization of the lemniscate pulse shape should be modified 
to keep the conditions \eqref{eq:lemniscate_gate_conditions}, this does not
affect our conclusions about the typical values of the infidelity and 
the scaling with $\eta$.

\section{Conclusions}
We analyze the role of out-of-Lamb-Dicke effects in GHZ state 
preparation in trapped-ion chains using perturbation theory in 
the Lamb-Dicke parameter $\eta$ and numerical simulations. 
We show that the out-of-Lamb-Dicke contribution 
into GHZ infidelity  
can be split into two parts: one part relates to the modification of the evolution operator in the qubit subspace, and another part relates to the phonon creation. Both of them grow fast with the increasing number of 
ions, and their combined contribution can reach $\sim 10^{-2}$ for tens of ions.
To mitigate these types of error, we propose an amplitude and phase-modulated 
laser pulse shape which allows canceling the leading-order 
contribution of the high-order terms. The error mitigation is 
achieved due to the special shape of the phase trajectory of the 
collective motional mode: the phase trajectory
is an figure-eight curve in the phase space. The residual error 
caused by higher-order terms in the Lamb-Dicke expansion scales 
as $\eta^6$ for the proposed pulse in contrast to $\eta^4$ 
for rectangular and bell-like pulses, 
and it stays significantly lower by the absolute value.
We show that our 
approach allows achieving the GHZ state infidelities below 
$10^{-4}$ for $20$-ion chains. The suggested pulse shapes can 
be readily implemented using pulse control systems of the 
state-of-the art ion traps, so they could become a 
valuable tool for high-fidelity quantum state engineering 
with cold trapped ions.






%

\begin{acknowledgments}
The work was supported by Rosatom in the framework of the Roadmap for Quantum
computing (Contract No. 868/1759-D dated 3 October 2025).
\end{acknowledgments}

\section*{Data availability}

The data that support the findings of this article are openly available \cite{LemniscateZenodo}.

\bibliography{references}

\appendix

\section{Ion chains stability conditions and the COM mode Lamb-Dicke parameter}
\label{appendix:ion_chains}
The equilibrium positions $\vec{r}_k = (x_k, y_k, z_k)$ of the ions 
in the three-dimensional harmonic potential are defined by the minimization 
of the total potential energy
\begin{equation}
U = \sum_k \frac{M(\omega_{x}^2x_k^2 + \omega_y^2 y_k^2 + \omega_z^2 z_k^2)}{2}
+ \sum_{k<l} \frac{e^2}{4\pi\epsilon_0|\vec{r}_k - \vec{r}_l|}.
\end{equation}
To ensure the stability of the linear configuration along 
the $z$-axis ($x_k = 0$, $y_k=0$), the radial frequencies $\omega_x$ 
and $\omega_y$ should be chosen large enough in comparison to $\omega_z$. The 
stability condition takes the form $\omega_{x,y}/\omega_z > a_n$,  where 
$a_n$ is the number-of-ions-dependent critical anisotropy. 
The critical anisotropy scales approximately as 
$\frac{3n}{4\sqrt{\log{n}}}$ \cite{Morigi2004}. Also, the trap secular frequencies 
usually do not exceed the values of several MHz, so the stability condition for longer
chains is achieved by lowering the axial frequency:
\begin{equation}
  \omega_z < \frac{\omega_{x,y}}{a_n} \approx \frac{4\sqrt{\log{n}}}{3n}\omega_{x,y}.
  \label{eq:stability_condition}
\end{equation}
The Lamb-Dicke parameter for the axial center-of-mass (COM) mode 
can be calculated from Eq.~\eqref{eq:ld_param_def} using the displacement
vector of the COM mode, $b_{i0} = 1/\sqrt{n}$:
\begin{equation}
  \eta = \sqrt{\frac{\omega_{rec}}{\omega_{z}n}},
  \label{eq:eta_com}
\end{equation}
where $\omega_{rec} = \frac{\hbar k_z^2}{2M}$ is the recoil frequency, 
$k_z$ is the laser field wavevector, and
$M$ is the ion mass. The Lamb-Dicke parameter depends on both $\omega_{z}$ 
and the number of ions. 
By combining Eqs.~\eqref{eq:stability_condition} and 
\eqref{eq:eta_com}, we obtain 
a very weak scaling of the Lamb-Dicke parameter with $n$:
\begin{equation}
\eta \sim 
\sqrt{ \frac{3\omega_{rec}}{4\omega_{x,y}}}(\log{n})^{-\frac14}.
\label{eq:ld_param_est}
\end{equation}
For optical qubits based on ${}^{40}\mathrm{Ca}^+$ ions, 
the recoil frequency is $\omega_{rec} = (2\pi)9390.6$ Hz. For the 
radial frequencies of $3\text{--}5$ MHz and the $n=2\text{--}20$, the 
values of the Lamb-Dicke parameter estimated from \eqref{eq:ld_param_est}
range from $0.03$ to $0.05$.

\section{The RWA Hamiltonian with account for all orders in Lamb-Dicke expansion}
\label{appendix:all_ord_rwa_ham}
Here we present the derivation of the effective RWA Hamiltonian 
\eqref{eq:all_ord_rwa_ham} from the full Hamiltonian \eqref{eq:general_ham}.
First, as discussed in Section~\ref{sec:setup_description}, we imply the 
single-mode approximation, so 
\begin{equation}
    k_z \hat{z} = \eta(\hat{a}e^{-i\omega_0 t} + \hat a^\dagger e^{i\omega_0 t}).
\end{equation}
In phase-sensitive geometry ($k_{z1} = k_{z2}$), the 
Hamiltonian \eqref{eq:general_ham} takes the form
\begin{equation}
  \hat{H} = -i\Omega(t)\cos{\mu t} 
  (\bigexp{i\eta(\hat{a}e^{-i\omega_0 t} + \hat a^\dagger e^{i\omega_0 t})}
  S_+ - \mathrm{h.c.}),
  \label{eq:single_mode_ps_ham}
\end{equation}
where $S_+ = \sum_{i} \sigma_+^{(i)}$.
In phase-insensitive geometry ($k_{z1} = -k_{z2}$), 
it reduces to
\begin{equation}
  \hat{H} = -i\Omega(t)\left(
    e^{-i\mu t}\bigexp{ 
     i\eta( \hat a e^{-i\omega_0 t} + \hat a^\dagger e^{i\omega_0 t})
    }
    - \mathrm{h.c.}
  \right)S_y.
  \label{eq:single_mode_pi_ham}
\end{equation}
To perform the rotating-wave approximation, let us represent the
exponent $\bigexp{i\eta( \hat a e^{-i\omega_0 t} + \hat a^\dagger e^{i\omega_0 t})}$ as a sum over its matrix elements:
\begin{multline}
    \bigexp{i\eta( \hat a e^{-i\omega_0 t} + \hat a^\dagger e^{i\omega_0 t})} \\=
    e^{ i\omega_0 t \hat a^\dagger \hat a} e^{i\eta(\hat a + \hat a^\dagger)}
    e^{-i\omega_0 t \hat a^\dagger \hat a}
    \\= \sum_{nn'} \langle n |e^{i\eta(\hat a + \hat a^\dagger)}|n'\rangle
    e^{i(n-n')\omega_0 t}|n\rangle\langle n'|.
\end{multline}
By substituting this expression into \eqref{eq:single_mode_ps_ham} and 
keeping only the terms oscillating with the frequencies 
$\pm \delta$, where $\delta = \mu - \omega_0$, one gets the approximate 
Hamiltonian \eqref{eq:all_ord_rwa_ham}. For \eqref{eq:single_mode_pi_ham},
the resulting Hamiltonian contains $\hat S_y$ instead of $\hat S_x$.

The matrix elements 
$\langle n |e^{i\eta(\hat a + \hat a^\dagger)}|n+1\rangle$
can be expressed through the generalized Laguerre polynomials
\cite{Soerensen2000}:
\begin{equation}
    \langle n |e^{i\eta(\hat a + \hat a^\dagger)}|n+1\rangle = 
    \frac{i\eta e^{-\eta^2/2}}{\sqrt{n+1}} L^1_n(\eta^2).
\end{equation}

\section{Leading-order correction to the evolution operator}
\label{appendix:magnus}
Here we present the derivation of the 
interaction-picture evolution operator
\eqref{eq:magnus_3ord_op} and analyze 
its contribution into GHZ preparation error. 

With the interaction-picture Hamiltonian
\eqref{eq:int_ham}, 
the leading-order $\hat T$ matrix takes the form
\begin{multline}
  \hat{T} = \int \hat{V}_I(t) dt
  = \sigma \hat a^\dagger \hat a^2 \hat S_x + \sigma^* (\hat a^\dagger)^2\hat a \hat S_x\\
  + h\hat a^\dagger \hat a \hat S_x^2
  + g_2^* \hat a^2 \hat S_x^2 + g_2 (\hat a^\dagger)^2 \hat S_x^2\\
  + g \hat a^\dagger \hat S_x^3 + g^* \hat a \hat S_x^3 
  + \theta_4 \hat \hat S_x^4,
  \label{eq:T_full}
\end{multline}
%
%
%

The coefficients $\sigma$, $h$, $g_2$, $g$ and $\theta_4$ can be found by 
integrating $\hat{V}$ given by Eq.~\eqref{eq:int_ham}. Also, it is convenient 
to express them only in terms of the phase trajectories $\alpha(t)$ using 
Eq.~\eqref{eq:alpha_sdf}:
\begin{equation}
  d\alpha = -\frac{i\eta}{2}\Omega(t) e^{-i\delta t}.
\end{equation}
Simple algebraic transformations lead to the 
following expressions:
\begin{equation}
   \sigma = i\eta^2\int d\alpha^*,
   \label{eq:sigma}
\end{equation}
\begin{equation}
   h = -2i\eta^2\int \alpha^* d\alpha - \alpha d\alpha^* = 2\eta^2\chi,
\end{equation}
\begin{equation}
   g_2 = 2i\eta^2\int \alpha^* d\alpha^*,
   \label{eq:g2}
\end{equation}
and $g$ and $\theta_4$ are given by 
Eqs.~\eqref{eq:ph_ex_ampl}, \eqref{eq:theta_4}.

For all laser pulses for GHZ state preparation
considered in the main text, phase trajectories
$\alpha(t)$ follow closed curves. Because of that,
the integrals \eqref{eq:sigma} and \eqref{eq:g2}
vanish. Furthermore, in the main text 
we assume that the phonon mode is ground-state-cooled
in the beginning of the GHZ preparation process.
Therefore, the term 
$h \hat a^\dagger \hat a \hat S_x^2$ does not 
contribute to the gate error. Because of that,
only the last three terms of the Eq.~\eqref{eq:T_full} contribute to the GHZ preparation 
infidelity.

\section{Numerical simulation of the GHZ preparation infidelity}
\label{appendix:numerical_simulation}
In Sections~\ref{sec:high_ord_effects}, \ref{sec:lemn_pulse}, we use the 
results of the numerical solution of the TDSE with the 
Hamiltonian~\eqref{eq:all_ord_rwa_ham}. The initial state is 
$|1\dots 1\rangle$ in the $z$-basis. As the Hamiltonian~\eqref{eq:all_ord_rwa_ham}
commutes with $\hat S_x$, it is convenient to decompose the initial
qubit state as a superposition of $\hat{S}_x$ eigenstates. Then,
the TDSE can be solved separately in each eigenspace of $S_x$, which have
considerably lower dimension than the full system Hilbert space. 
The decomposition of $|1\dots 1\rangle$ takes the following form:
\begin{equation}
  |1\dots 1\rangle_z = |n/2,n/2\rangle_z 
  = \frac{1}{2^{n/2}} \sum_m \sqrt{C_{n}^{n/2-m}}  |n/2, m\rangle_x.
\end{equation}
For the initial states $|n/2,m\rangle_x\otimes |0\rangle$, the 
result of the TDSE solution takes the form 
$|n/2,m\rangle_x\otimes |\psi_m\rangle$, where $|\psi_m\rangle$ belongs to the 
phonon Hilbert space.
Then, the full final state can be found as the following superposition:
\begin{equation}
  |\psi_{fin}\rangle = \frac{1}{2^{n/2}}\sum_m  
  \sqrt{C_{n}^{n/2-m}}  |n/2, m\rangle_x\otimes |\psi_m\rangle.
\end{equation}
For the ideal global MS gate operation $\hat{U} = e^{-\frac{i\pi}{2}\hat{S}_x^2}$,
$|\psi_m\rangle = e^{-\frac{i\pi m^2}{2}}|0\rangle$.
Therefore, GHZ preparation fidelity can be calculated with the equation
\begin{equation}
  F = |\langle\psi_{id} | \psi_{fin}\rangle |^2 =
    \frac{1}{2^{2n}}
    \left|
    \sum_m  e^{\frac{i\pi m^2}{2}} C_{n}^{n/2-m} \langle 0 | \psi_m\rangle 
    \right|^2.
\end{equation}

\end{document}